\documentclass[12pt,preprint]{aastex}

\shorttitle{Swift observations of GRB050525a}
\shortauthors{A. J. Blustin et al.}
\begin{document}
\title{Swift panchromatic observations of the bright gamma-ray burst GRB050525a}
\author{A. J. Blustin$^1$, D. Band$^2$, S.~Barthelmy$^2$, P. Boyd$^2$, M. Capalbi$^4$, 
S. T. Holland$^2$, F.~E.~Marshall$^2$, K.~O.~Mason$^1$, M.~Perri$^4$, T. Poole$^1$, 
P. Roming$^3$, S. Rosen$^1$, P. Schady$^{1,3}$, M. Still$^2$, B. Zhang$^5$,
L. Angelini$^2$, L. Barbier$^2$, A. Beardmore$^6$,  A. Breeveld$^1$, D. N. Burrows$^3$, 
J.~R.~Cummings$^2$, J. Cannizzo$^2$, 
S. Campana$^8$, M. M. Chester$^3$,  G.  Chincarini$^8$, 
L.R. Cominsky$^9$,  A. Cucchiara$^3$, M. de Pasquale$^1$, E.E. Fenimore$^7$, N. Gehrels$^2$, P. Giommi$^4$,
M. Goad$^6$,  C.~Gronwall$^3$, D. Grupe$^3$, J.E. Hill$^2$, D. Hinshaw$^2$,  S. Hunsberger$^3$, K.C. Hurley$^{11}$, 
M.~Ivanushkina$^3$, J.A. Kennea$^3$,  H. A. Krimm$^2$, P. Kumar, W. Landsman$^2$, V. La Parola$^{10}$, C.~B.~Markwardt$^2$,  
K. McGowan$^1$, P. M\'esz\'aros$^3$,  T. Mineo$^{10}$,  A. Moretti$^8$, A.~Morgan$^3$,  
J. Nousek$^3$,  P.T. O'Brien$^6$, J.P. Osborne$^6$, K. Page$^6$, M. J. Page$^1$,  D. M. Palmer$^7$, 
A.M. Parsons$^2$, J.~Rhoads$^{13}$, P.~Romano$^8$, T.  Sakamoto$^2$, 
G. Sato$^{12}$, G. Tagliaferri$^8$,  J. Tueller$^2$, A.A. Wells$^6$ 
\&
N.~E.~White$^2$ 
} 
\affil{$^1$Mullard Space Science Laboratory, Department of Space and Climate
Physics, \\ University College London, Holmbury St Mary, Dorking, Surrey,
RH5 6NT, UK.\\
$^2$ NASA/Goddard Space Flight Center, Greenbelt, MD 20771, USA.\\
$^3$ Department of Astronomy and Astrophysics, Pennsylvania State University, 525 Davey Laboratory, University Park, PA 16802, USA.\\
$^4$ ASI Science Data Center, Via Galileo Galilei, I-00044 Frascati, Italy.\\
$^5$ Department of Physics, University of Nevada, Las Vegas, Nv 89154, USA.\\
$^6$ Department of Physics \& Astronomy, University of Leicester, Leicester LE1 7RH, UK.\\
$^7$ Los Alamos National Laboratory, Los Alamos, New Mexico 87545, USA.\\
$^8$ INAF - Osservatorio Astronomico di Brera, Via Bianchi 46, I-23807 Merate, Italy.\\
$^{10}$ INAF - Istituto di Astrofisica Spaziale e Cosmica, Via Ugo La Malfa 153, I-90146 Palermo, Italy.\\
$^{11}$ UC Berkeley Space Sciences Laboratory, Berkeley, Ca 94720, USA.\\
$^{12}$ Institute of Space and Astronautical Science, Kanagawa 229-8510, Japan.\\
$^{13}$ Space Telescope Science Institute, 3700 San Martin Drive, Baltimore, MD 21218, USA.\\
} 
\email{ajb@mssl.ucl.ac.uk}
\received{2005 May 26}
\begin{abstract}
The bright gamma-ray burst GRB050525a has been detected with the Swift 
observatory, providing unique multiwavelength coverage from the very earliest phases of the burst. 
The X-ray and optical/UV afterglow decay light curves both exhibit a steeper slope $\sim 0.15$ days after the burst, 
indicative of a jet break. This jet break time combined with the total gamma-ray energy of the burst
constrains the opening angle of the jet
to be $3.2^\circ$. We derive an empirical `time-lag' redshift from the BAT data of $\hat z = 0.69 \pm 0.02$, 
in good agreement with the spectroscopic redshift of 0.61.

Prior to the jet break, the X-ray data can be modelled by a simple power law with index
$\alpha = -1.2$. However after 300~s the X-ray flux brightens by about 30\%\ compared to the power-law fit. 
The optical/UV data have a  more complex decay, with evidence of a 
rapidly falling reverse shock component that
dominates in the first minute or so, giving way to a flatter forward shock component at later times. 
The multiwavelength X-ray/UV/Optical
spectrum of the afterglow shows evidence for migration of the electron cooling frequency through the optical range
within 25000s. The measured temporal decay 
and spectral indices in the X-ray and optical/UV regimes compare favourably with the standard fireball model 
for Gamma-ray bursts assuming expansion into a constant density interstellar medium. 
\end{abstract}


\clearpage

\section{Introduction}

The observation of Gamma-ray bursts has entered a new era with the launch of the Swift observatory 
(Gehrels et al. 2004). Swift provides rapid notification of GRB triggers to the ground using its sensitive 
Burst Alert Telescope (BAT; Barthelmy et al. 2005), and can make panchromatic observations of the burst and 
its afterglow by bringing its narrow field X-ray Telescope (XRT; Burrows et al. 2005a) and UltraViolet/Optical 
Telescope (UVOT; Roming et al. 2005a) to bear within about 1 minute of the burst going off. 

In this paper we 
describe Swift observations of GRB050525a, the first bright, low-redshift burst to have been observed with Swift 
with both its narrow field instruments. This has resulted in the most complete multiwavelength description of the early 
evolution of a burst afterglow yet obtained.
We describe the light curve decay in the X-ray and optical/UV bands, and compare
these with the predictions of theoretical models. We also consider how the multiwavelength spectrum of the burst afterglow 
evolves with time, making use of the simultaneous X-ray/UV/optical coverage. We combine the measurements of the 
afterglow properties with the well constrained measurements of the prompt Gamma-ray emission to constrain the geometry
of the burst emission. We present an analysis of the prompt gamma-ray emission in section 2.1. This is followed by an 
analysis of
the time evolution and spectrum of the X-ray data in section 2.2, adding to this the optical and 
ultraviolet data in section 2.3
to build up the multiwavelength picture. The results are discussed in section 3.

\section{Observations and Analysis}

\subsection{BAT}

At 00:02:53.26 UT, the {Swift} Burst Alert Telescope
(BAT) triggered and located on-board GRB050525a (trigger
\#130088).  The BAT location calculated on-board was RA,
DEC 278.144, +26.340 (J2000) with an uncertainty of
4~arcmin (radius, 3$\sigma$, including estimated systematic
uncertainty). This was later refined (see Figure~1) to 278.140, +26.344
$\pm$ 0.5' (95\% containment, the position corresponding to 18h 32m 34s, +26d 20' 38'')
using the full data set downloaded through the Malindi
groundstation $\sim$1 hour after the burst. The refined
position is 26 arcsec from the on-board location and 21
arcsec from the UVOT location. The burst was 28$\arcdeg$
off the BAT bore sight (85\% coded) and was 148$\sigma$ in
the image-domain full data set. The spacecraft completed
the automated slew to the burst at 00:04:58 UT (75 sec
after the trigger), and XRT and UVOT began their
standard follow-up observation sequences.

Figure~2 shows the BAT light curves in the 4 standard BAT
energy bands. The burst was very bright and
was detected in all 4 energy bands with a peak rate of
$\sim$80,000~cts~s$^{-1}$. The burst has two main peaks with several smaller
peaks within the main structure. 
The lag between
the lightcurves in the 100--350 keV and the 25--50~keV
bands is $\tau=0.124\pm0.006$~s (the high energy band leads
the low energy band).  
The total duration was $\sim$12~s
($T_{90}$ is 8.8$\pm$0.5~s).  The successful trigger criterion
for this burst was in the 25--100~keV band with a duration
of 64~msec.  

Another very bright burst occurred
$\sim$2800~s later (GRB050525b; Golenetskii et al. 2005). This was outside the
field-of-view (FOV) of INTEGRAL's IBIS instrument whereas
GRB050525a was inside the IBIS FOV. Thus we can be confident that this
second burst was not associated with GRB050525a
(Mereghetti et al. 2005).

We fit BAT spectra using the current analysis techniques
developed by the instrument team.  Systematic effects are
still being resolved this early in the mission, and
consequently the fits described here should be regarded as
preliminary. To reproduce the known spectrum of the Crab,
empirical correction factors have been applied to the
pre-launch response matrix; this analysis also resulted in
an estimate of the energy-dependent systematic error that
should be added to the statistical error.  The corrections
and the systematic error are posted.\footnote{
http://legacy.gsfc.nasa.gov/docs/swift/analysis/bat\_digest.html}
The count spectra are binned between 16 and 148.8~keV in
$\sim$2~keV bins.  Table~1 summarizes the spectral fits to
the entire burst (12.80~s) and to the peak 1~s, with 90\%
confidence limits. We fit the count spectra with three
nested models, here presented as energy spectra\footnote{We
use the notation $\beta$ here and throughout the paper to
denote the power law index of the energy spectrum of the
source.}: a power law $F(E)\propto E^{\beta}$; a power law
with an exponential cutoff $F(E)\propto
E^{\beta}\exp[-E/E_0]$; and the `Band' model (Band et al.
1993), a low energy power law with an exponential cutoff
that transitions into a high energy power law $F(E)\propto
E^{\beta _2}$.  The peak energy $E_p=(1+\beta)E_0$ is both
physically more relevant and less correlated with $\beta$
than $E_0$; $E_p$ is the energy of the peak of $E\
F(E)\propto \nu f_\nu$.  The power law with an exponential
cutoff is the same as the Band model with
$\beta_2=-\infty$, and the power law model is the same as
the other two models with $E_0=\infty$.



\begin{deluxetable}{l l c c c c c c c r}
\tablecolumns{10}
\tabletypesize{\footnotesize}
\tablewidth{0in}
\tablecaption{\label{Table1}BAT Spectral Fits}
\tablehead{
& \vline & \multispan3{\hfil  Entire Burst \hfil }  & \vline & \multispan3{\hfil Peak Flux \hfil} & \vline\\
\colhead{Parameter}& \vline &
\colhead{Power Law\tablenotemark{a}} &
\colhead{Power Law,} &
\colhead{Band Model\tablenotemark{c}} & \vline& 
\colhead{Power Law\tablenotemark{a}} &
\colhead{Power Law,} &
\colhead{Band Model\tablenotemark{c}} & \vline\\
&\vline& &Cutoff\tablenotemark{b}&& \vline & &Cutoff\tablenotemark{b}& &\vline}
\startdata
$\beta$\tablenotemark{d} &    & 
$-0.78$\tablenotemark{e} & $0.01^{+0.11}_{-0.12}$ & $0.01^{+0.11}_{-0.12}$ &    & 
-0.42\tablenotemark{e} & $0.45^{+0.14}_{-0.14}$ & $-0.45^{+0.14}_{-0.14}$ &    \\
$\beta _2$\tablenotemark{f} &    & 
 --- & --- & $-7.84^{+6.40}_{-1.16}$ &   &  --- & --- & $-8.27^{+7.14}_{-0.73}$ &    \\
$E_p$\tablenotemark{g} &    & 
---& $78.8^{+3.9}_{-3.1}$ & $78.8^{+3.7}_{-3.1}$ &   
& --- & $102.4^{+7.1}_{-6.3}$ & $102.4^{+8.1}_{-6.3}$ &    \\
Norm &    & 
7.15\tablenotemark{e,h} & $14.2^{+5.9}_{-4.2}$\ \tablenotemark{i} & $15.0^{+1.65}_{-1.45}$\ \tablenotemark{h} &   
& 19.35\tablenotemark{e,h} & $7.62^{+3.91}_{-2.65}$\ \tablenotemark{i} & $44.00^{+6.05}_{-5.20}$\ \tablenotemark{h} &    \\

$\chi^2$/dof &    & 
181.7/57 & 15.2/56 & 15.2/55  &   
& 169.8/57 & 31.6/56 & 31.6/55  &    \
\enddata
\tablenotetext{a}{Power law model, $F(E)\propto E^{\beta}$.}
\tablenotetext{b}{Power law with an exponential cutoff,
$F(E)\propto E^{\beta}\exp[-E/E_0]$.}
\tablenotetext{c}{Band model (Band et al. 1993), a low energy power
law with an exponential cutoff transitioning to a high energy
power law $F(E)\propto E^{\beta _2}$.}
\tablenotetext{d}{The low energy spectral index.}
\tablenotetext{e}{Fit too poor to produce uncertainty range.}
\tablenotetext{f}{The high energy spectral index. The fit is insensitive
to $\beta_2<-2.5$ for the fitted $E_p$.}
\tablenotetext{g}{The energy of the peak of $EN(E)\propto \nu f_\nu$,
and $E_p=(2+\beta)E_0$.}
\tablenotetext{h}{The normalization of the spectrum at 50~keV, in
keV cm$^{-2}$ s$^{-1}$ keV$^{-1}$.}
\tablenotetext{i}{The normalization of the spectrum at 1~keV, in
keV cm$^{-2}$ s$^{-1}$ keV$^{-1}$.}

\end{deluxetable}

As can be seen from these fits, the power law fit is
definitely inferior to the other two spectral models, and
the second and third spectral models are equivalent in this case since the
transition to the high energy power law at $E_b =
(\beta-\beta_2)E_0 = (\beta-\beta_2)E_p/(\beta+1)$ is above
150~keV (the energy of the highest PHA channel that is fitted);
indeed, for the observed $\beta$ and $E_p$ values, the fits
are insensitive to $\beta_2<-2.5$.


Konus-Wind (Golenetskii et al. 2005) and {\it INTEGRAL}
(Gotz et al. 2005) both detected GRB050525a.  Golenetskii et al. (2005) fit the Konus-Wind
20~keV to 1~MeV spectrum for the entire burst with a
power law with an exponential cutoff model and find
$\beta = -0.10 \pm 0.05$, and $E_p = 84.1 \pm 1.7$~keV,
consistent with the {\it Swift}
fit.

The spectral fits can be integrated over energy and time
to give a fluence $S(\hbox{15--350 keV})$ =
$(2.01\pm0.05)\times 10^{-5}$ erg cm$^{-2}$ and a peak flux
of $P=(\hbox{15--350 keV}) = 47.7 \pm 1.2$ photons
cm$^{-2}$ s$^{-1}$.  Using the observed redshift of
$z=0.606$ (Foley et al. 2005), the spectral fits and a
standard cosmology ($\Omega_m$=0.3, $\Omega_\Lambda$=0.7,
and $H_0=70$~km~s$^{-1}$~Mpc$^{-1}$), we find $E_{\rm
iso}=2.3\times 10^{52}$~erg and $L_{\rm iso} = 7.8\times
10^{51}$~ erg~s$^{-1}$.  These two derived quantities are
the total energy and peak luminosity that would have been
radiated if the observed flux were emitted
isotropically. As will be discussed below (section 3) in considering 
the possible jet break,  
the emission was most likely
beamed and the actual emitted energy and luminosity were
consequently smaller.  Using the redshift, the peak
energies in the burst frame are $E_{\rm
pp}=164.5\pm11.6$~keV for the peak of the lightcurve and
$E_{\rm pt}=126.6\pm5.5$~keV for the entire burst.

\subsection{XRT data}
The XRT observations started on 2005 May 25 at 00:04:08 UT, 75 seconds 
after the trigger. The XRT was undergoing engineering tests at the time
of the burst and the Auto State (which selects the observing mode based on source brightness) 
was disabled. A sequence of 3 frames in 
Photon Counting (PC) mode were taken first, between T+75s to T+83s.
These early PC mode data suffer from severe pile-up because of the brightness
of the source, and have been excluded from the analysis. 

Shortly afterward the instrument was put into Auto State and a 2.5~s
exposure in Image mode was taken starting at 00:04:58 UT, 125 seconds 
after the BAT burst trigger. A bright X--ray source was found near the
center of the field of view. The refined X--ray coordinates are 
RA(J2000) = $18^{\mathrm h}32^{\mathrm m}32^{\mathrm s}.6$ and 
Dec(J2000) = $26^{\circ}20'18''$, with an estimated positional 
uncertainty of 6 arc seconds (90\% confidence level).
The XRT coordinates are 4 arc seconds from the UVOT position of the
optical counterpart (see below).

Following the automatic sequence of readout modes, designed to avoid any
pile-up effect, the instrument was configured in the Photodiode (PD) 
readout mode starting from 00:05:01 UT (T+128~s). Due to the engineering
tests mentioned above, the instrument remained in PD mode 
until 00:20:21 UT (T+1048~s) even when the afterglow brightness was below the
nominal flux threshold used for this mode. 

Data in PC mode, the most sensitive XRT operational mode,
were taken starting at 01:40:32 UT (T+5859~s) until 11:39:11 UT 
(T+41778~s).
The GRB050525a field was re-observed in PC mode on several occasions up to late June 2005.

\subsubsection{Temporal analysis}

For the PD mode, events for the temporal analysis were selected in the 
0.4--4.5 keV energy band to avoid contamination from the calibration 
sources. The standard grade selection for this mode (0--5) was used. The 
count rate was converted to unabsorbed 2--10 keV flux using the  
best fit spectral model (see below). 

For PC data, events were selected in the 0.3--10 keV band and grades 0--12 were used 
in the analysis. Furthermore, since 2-D spatial information is available for 
this mode, photons within a circle of 10 pixel ($\sim$24 arcsec) radius, 
which encloses about 80\%
of the PSF at 1.5 keV, were extracted, centered on the source position. The 
background was estimated from a nearby source-free circular region with
50 pixel radius. Again, the count rate in the 0.3--10 keV band was 
converted to unabsorbed 2--10 keV flux using the best fit spectral model.

\begin{deluxetable}{r r r r r r r}
\tablecolumns{7}
\tabletypesize{\footnotesize}
\tablewidth{0in}
\tablecaption{\label{Table2}XRT 2-10 keV flux}
\tablehead{
\colhead{T(mid)\tablenotemark{a}} &
\colhead{T(exp)} &
\colhead{\hfil Flux\tablenotemark{b}} & 
\colhead{   } & 
\colhead{T(mid)\tablenotemark{a}} &
\colhead{T(exp)} &
\colhead{\hfil Flux\tablenotemark{b}}  \\
 (s) \phm{a}\hfill & \hfill (s)\phm{a}\hfill & & \phm{aaa} &\hfill (s)\phm{a}\hfill  & \hfill (s)\phm{a}\hfill \\
}
\startdata
133   &     5. &    122.7 $\pm$      5.7  && 578   &    10. &     29.8 $\pm$      2.0  \\ 
143   &     5. &    109.5 $\pm$      5.4  && 598   &    10. &     28.5 $\pm$      2.0  \\ 
153   &     5. &    101.4 $\pm$      5.2  && 618   &    10. &     29.1 $\pm$      2.0  \\ 
163   &     5. &     92.0 $\pm$      4.9  && 638   &    10. &     24.8 $\pm$      1.8  \\ 
173   &     5. &     86.8 $\pm$      4.8  && 658   &    10. &     27.3 $\pm$      1.9  \\ 
183   &     5. &     83.7 $\pm$      4.7  && 678   &    10. &     24.6 $\pm$      1.8  \\ 
193   &     5. &     77.2 $\pm$      4.5  && 708   &    20. &     24.2 $\pm$      1.3  \\ 
203   &     5. &     69.4 $\pm$      4.3  && 748   &    20. &     20.4 $\pm$      1.2  \\ 
213   &     5. &     69.2 $\pm$      4.3  && 788   &    20. &     19.8 $\pm$      1.2  \\ 
223   &     5. &     62.4 $\pm$      4.1  && 828   &    20. &     19.0 $\pm$      1.1  \\ 
233   &     5. &     65.0 $\pm$      4.1  && 868   &    20. &     16.3 $\pm$      1.1  \\ 
243   &     5. &     57.2 $\pm$      3.9  && 908   &    20. &     18.5 $\pm$      1.1  \\ 
253   &     5. &     54.6 $\pm$      3.8  && 948   &    20. &     17.6 $\pm$      1.1  \\ 
263   &     5. &     54.3 $\pm$      3.8  && 988   &    20. &     16.3 $\pm$      1.1  \\ 
278   &    10. &     50.8 $\pm$      2.6  && 1028   &    20. &     15.5 $\pm$      1.4  \\ 
298   &    10. &     49.8 $\pm$      2.6  && 6009   &   150. &    1.072 $\pm$    0.134  \\ 
318   &    10. &     45.4 $\pm$      2.5  && 6309   &   150. &    1.331 $\pm$    0.153  \\ 
338   &    10. &     45.5 $\pm$      2.5  && 6609   &   150. &    0.987 $\pm$    0.126  \\ 
358   &    10. &     46.9 $\pm$      2.5  && 6909   &   150. &    1.010 $\pm$    0.152  \\ 
378   &    10. &     40.2 $\pm$      2.3  && 11809   &   350. &    0.560 $\pm$    0.056  \\ 
398   &    10. &     41.7 $\pm$      2.4  && 12509   &   350. &    0.482 $\pm$    0.052  \\ 
418   &    10. &     39.4 $\pm$      2.3  && 18109   &   350. &    0.331 $\pm$    0.052  \\ 
438   &    10. &     39.9 $\pm$      2.3  && 23859   &  2000. &    0.173 $\pm$    0.039  \\ 
458   &    10. &     34.8 $\pm$      2.2  && 27859   &  2000. &    0.108 $\pm$    0.028  \\ 
478   &    10. &     31.9 $\pm$      2.1  && 35859   &  2000. &    0.079 $\pm$    0.021  \\ 
498   &    10. &     31.5 $\pm$      2.1  && 81459   &  5400. &   0.0361 $\pm$   0.0088  \\ 
518   &    10. &     32.5 $\pm$      2.1  && 99102   & 11850. &   0.0273 $\pm$   0.0078  \\ 
538   &    10. &     27.4 $\pm$      1.9  && 165485   & 22000. &   0.0080 $\pm$   0.0016  \\ 
558   &    10. &     32.4 $\pm$      2.1  && 412515   & 42000. &   0.0013 $\pm$   0.0006  \\ 
\enddata
\tablenotetext{a}{time since trigger}
\tablenotetext{b}{flux in units of $10^{-11}$ erg cm$^{-2}$ s$^{-1}$ 2-10 keV}
\end{deluxetable}

\begin{deluxetable}{r r r r r r r r r  }
\tablecolumns{9}
\tabletypesize{\footnotesize}
\tablewidth{0in}
\tablecaption{\label{Table3}UVOT multicolor data}
\tablehead{
\colhead{T(mid)\tablenotemark{a}} &
\colhead{T(exp)} &
\colhead{Mag} &
\colhead{Flux Density\tablenotemark{b}}  & &
\colhead{T(mid)\tablenotemark{a}} &
\colhead{T(exp)} &
\colhead{Mag} &
\colhead{Flux Density\tablenotemark{b}}  \\
(s) \phm{a}   \hfill & \hfill (s)\phm{a}   \hfill & & & \phm{aaa}&(s) \phm{a}   \hfill & \hfill (s)\phm{a}   \hfill \\
}
\startdata
\sidehead{V filter} 
66    &     1. &  13.21 $\pm$   0.24 &    189.8 $\pm$     41.2  && 148    &     5. &  13.84 $\pm$   0.12 &    106.2 $\pm$     11.5  \\ 
67    &     1. &  12.90 $\pm$   0.23 &    254.0    $\pm$     53.6  && 153    &     5. &  13.87 $\pm$   0.12 &    103.2 $\pm$     11.3  \\ 
68    &     1. &  12.86 $\pm$   0.23 &    263.3 $\pm$     55.5  && 158    &     5. &  14.06 $\pm$   0.12 &     87.1 $\pm$     10.0  \\ 
69    &     1. &  13.01 $\pm$   0.23 &    227.9 $\pm$     48.5  && 163    &     5. &  14.00 $\pm$   0.12 &     91.7 $\pm$     10.3  \\ 
70    &     1. &  12.97 $\pm$   0.23 &    236.3 $\pm$     50.1  && 168    &     5. &  14.01 $\pm$   0.12 &     90.8 $\pm$     10.3  \\ 
71 &     1. &  13.31 $\pm$   0.23 &    172.8 $\pm$     34.0  && 173   &    10. &  14.08 $\pm$   0.13 &     83.3 $\pm$     10.6  \\ 
72   &     1. &  13.13 $\pm$   0.23 &    204.3 $\pm$     43.9  && 258   &    10. &  14.64 $\pm$   0.14 &     49.8 $\pm$      6.8  \\ 
73   &     1. &  13.01 $\pm$   0.23 &    227.9 $\pm$     48.5  && 342   &    10. &  14.79 $\pm$   0.15 &     43.3 $\pm$      6.4  \\ 
78   &     5. &  13.13 $\pm$   0.10 &    204.3 $\pm$     19.6  && 426   &    10. &  15.22 $\pm$   0.17 &     29.2 $\pm$      4.9  \\ 
83   &     5. &  13.26 $\pm$   0.10 &    181.5 $\pm$     17.7  && 511   &    10. &  15.47 $\pm$   0.19 &     23.2 $\pm$      4.4  \\ 
88   &     5. &  13.18 $\pm$   0.10 &    195.5 $\pm$     18.9  && 595   &    10. &  16.06 $\pm$   0.24 &     13.4 $\pm$      3.3  \\ 
93   &     5. &  13.24 $\pm$   0.11 &    185.6 $\pm$     18.1  && 680   &    10. &  15.83 $\pm$   0.22 &     16.6 $\pm$      3.7  \\ 
98   &     5. &  13.25 $\pm$   0.11 &    184.2 $\pm$     17.9  && 764   &    10. &  16.06 $\pm$   0.25 &     13.4 $\pm$      3.5  \\ 
103   &     5. &  13.51 $\pm$   0.11 &    144.9 $\pm$     14.7  && 849   &    10. &  15.78 $\pm$   0.22 &     17.4 $\pm$      3.9  \\ 
108   &     5. &  13.44 $\pm$   0.11 &    154.4 $\pm$     15.5  && 933   &    10. &  15.85 $\pm$   0.24 &     16.3 $\pm$      4.0  \\ 
113   &     5. &  13.67 $\pm$   0.11 &    124.7 $\pm$     13.0    && 1243   &   100. &  16.34 $\pm$   0.15 &     10.4 $\pm$      1.5  \\ 
118   &     5. &  13.48 $\pm$   0.11 &    148.4 $\pm$     15.0  && 18575   &   156. &  18.15 $\pm$   0.41 &      2.0 $\pm$      0.9  \\ 
123   &     5. &  13.62 $\pm$   0.11 &    130.2 $\pm$     13.5  && 22163   &   580. &  19.10 $\pm$   0.27 &      0.8 $\pm$      0.2  \\ 
128   &     5. &  13.86 $\pm$   0.12 &    104.2 $\pm$     11.4  && 35638   &   750. &  18.86 $\pm$   0.27 &      1.0 $\pm$      0.3  \\ 
133   &     5. &  13.70 $\pm$   0.11 &    121.5 $\pm$     12.8  && 49320   &  4982. & $>$ 20.62\hfil & $<$     0.2 \hfil \\ 
138   &     5. &  13.83 $\pm$   0.12 &    107.2 $\pm$     11.6  && 971360   & 33800. & $>$ 22.09\hfil & $<$     0.1 \hfil \\ 
143   &     5. &  13.81 $\pm$   0.12 &    109.2 $\pm$     11.8  && 1171176   &  6081. & $>$ 21.16\hfil & $<$     0.1 \hfil \\ 
\sidehead{B filter}
229   &    10. &  14.79 $\pm$   0.12 &     72.2 $\pm$      8.4  && 904   &    10. &  16.44 $\pm$   0.20 &     15.8 $\pm$      3.2  \\ 
313   &    10. &  15.19 $\pm$   0.12 &     49.9 $\pm$      5.8  && 1034   &   100. &  16.61 $\pm$   0.11 &     13.5 $\pm$      1.4  \\ 
397   &    10. &  15.51 $\pm$   0.13 &     37.2 $\pm$      4.7  && 12671   &   390. &  18.59 $\pm$   0.18 &      2.2 $\pm$      0.4  \\ 
482   &    10. &  15.63 $\pm$   0.14 &     33.3 $\pm$      4.6  && 16182   &   190. &  18.69 $\pm$   0.17 &      2.0 $\pm$      0.3  \\ 
571   &    10. &  15.70 $\pm$   0.14 &     31.2 $\pm$      4.3  && 30031   &   388. &  19.82 $\pm$   0.52 &      0.7 $\pm$      0.4  \\ 
651   &    10. &  16.13 $\pm$   0.16 &     21.0 $\pm$      3.3  && 33898   &   900. &  20.84 $\pm$   0.45 &      0.3 $\pm$      0.1  \\ 
735   &    10. &  16.03 $\pm$   0.16 &     23.0 $\pm$      3.7  && 45468   &   896. & $>$ 20.70\hfil & $<$     0.3 \hfil \\ 
820   &    10. &  16.56 $\pm$   0.20 &     14.1 $\pm$      2.9  && 62549   &  6513. & $>$ 21.55\hfil & $<$     0.1 \hfil \\ 
\sidehead{U filter}
215   &    10. &  13.70 $\pm$   0.18 &    110.3 $\pm$     19.9  && 890   &    10. &  15.29 $\pm$   0.21 &     25.5 $\pm$      5.4  \\ 
299   &    10. &  14.08 $\pm$   0.18 &     77.8 $\pm$     14.0  && 975   &    10. &  15.32 $\pm$   0.22 &     24.8 $\pm$      5.6  \\ 
419   &    10. &  14.47 $\pm$   0.19 &     54.3 $\pm$     10.4  && 12019   &   900. &  17.66 $\pm$   0.17 &      2.9 $\pm$      0.5  \\ 
468   &    10. &  14.71 $\pm$   0.19 &     43.5 $\pm$      8.3  && 28664   &   194. &  18.85 $\pm$   0.30 &      1.0 $\pm$      0.3  \\ 
552   &    10. &  14.97 $\pm$   0.20 &     34.3 $\pm$      6.9  && 29380   &   900. &  18.79 $\pm$   0.20 &      1.0 $\pm$      0.2  \\ 
637   &    10. &  15.01 $\pm$   0.20 &     33.0 $\pm$      6.7  && 41040   &   758. &  19.99 $\pm$   0.54 &      0.3 $\pm$      0.2  \\ 
721   &    10. &  15.14 $\pm$   0.20 &     29.3 $\pm$      5.9  && 52994   &   780. & $>$ 19.68\hfil & $<$     0.4 \hfil \\ 
805   &    10. &  15.46 $\pm$   0.21 &     21.8 $\pm$      4.7  && 128928   &  1548. & $>$ 20.05\hfil & $<$     0.3 \hfil \\ 
\sidehead{UVW1 (260 nm) filter }
202   &    10. &  13.61 $\pm$   0.07 &    162.8 $\pm$     10.4  && 877   &    10. &  16.15 $\pm$   0.24 &     15.7 $\pm$      3.5  \\ 
286   &    10. &  14.18 $\pm$   0.09 &     96.3 $\pm$      8.0  && 961   &    10. &  15.76 $\pm$   0.20 &     22.6 $\pm$      4.1  \\ 
370   &    10. &  14.77 $\pm$   0.12 &     56.2 $\pm$      6.2  && 6761   &   638. &  17.33 $\pm$   0.06 &      5.3 $\pm$      0.3  \\ 
455   &    10. &  15.10 $\pm$   0.14 &     41.5 $\pm$      5.3  && 28111   &   900. &  19.49 $\pm$   0.19 &      0.7 $\pm$      0.1  \\ 
539   &    10. &  15.22 $\pm$   0.15 &     37.0 $\pm$      5.1  && 40585   &   900. &  20.33 $\pm$   0.39 &      0.3 $\pm$      0.1  \\ 
624   &    10. &  15.26 $\pm$   0.15 &     35.6 $\pm$      4.9  && 52100   &   898. & $>$ 21.05\hfil & $<$     0.3 \hfil \\ 
708   &    10. &  15.22 $\pm$   0.15 &     36.9 $\pm$      5.1  && 63738   &   900. & $>$ 21.20\hfil & $<$     0.3 \hfil \\ 
792   &    10. &  15.88 $\pm$   0.21 &     20.1 $\pm$      3.8  &&   \\ 
\sidehead{UVM2 (220 nm) filter}
187   &    10. &  13.54 $\pm$   0.10 &    218.9 $\pm$     20.0  && 863   &    10. &  15.57 $\pm$   0.26 &     33.8 $\pm$      8.1  \\ 
272   &    10. &  14.32 $\pm$   0.14 &    106.8 $\pm$     14.1  && 947   &    10. &  15.66 $\pm$   0.27 &     31.2 $\pm$      7.6  \\ 
356   &    10. &  14.60 $\pm$   0.16 &     82.4 $\pm$     12.4  && 5984   &   900. &  17.10 $\pm$   0.06 &      8.2 $\pm$      0.4  \\ 
440   &    10. &  14.94 $\pm$   0.19 &     60.3 $\pm$     10.7  && 24007   &   864. &  18.96 $\pm$   0.19 &      1.5 $\pm$      0.3  \\ 
525   &    10. &  15.84 $\pm$   0.30 &     26.4 $\pm$      7.2  && 39683   &   897. &  19.72 $\pm$   0.30 &      0.7 $\pm$      0.2  \\ 
609   &    10. &  14.90 $\pm$   0.19 &     62.4 $\pm$     10.8  && 51257   &   900. &  20.34 $\pm$   0.48 &      0.4 $\pm$      0.2  \\ 
694   &    10. &  15.18 $\pm$   0.21 &     48.3 $\pm$      9.4  && 74798   &   931. & $>$ 21.59\hfil & $<$     0.6 \hfil \\ 
778   &    10. &  15.51 $\pm$   0.26 &     35.9 $\pm$      8.5  &&   \\ 
\sidehead{UVW2 (198 nm) filter} 
245   &    10. &  14.69 $\pm$   0.11 &     88.4 $\pm$      9.5  && 835   &    10. &  16.43 $\pm$   0.27 &     17.8 $\pm$      4.5  \\ 
329   &    10. &  15.30 $\pm$   0.16 &     50.2 $\pm$      7.2  && 920   &    10. &  16.21 $\pm$   0.25 &     21.8 $\pm$      4.9  \\ 
413   &    10. &  15.13 $\pm$   0.14 &     58.6 $\pm$      7.7  && 1140   &   100. &  16.35 $\pm$   0.08 &     19.1 $\pm$      1.5  \\ 
498   &    10. &  15.60 $\pm$   0.18 &     38.2 $\pm$      6.4  && 17973   &   882. &  18.99 $\pm$   0.13 &      1.7 $\pm$      0.2  \\ 
582   &    10. &  15.22 $\pm$   0.15 &     54.3 $\pm$      7.5  && 34806   &   900. &  20.05 $\pm$   0.25 &      0.6 $\pm$      0.1  \\ 
666   &    10. &  15.80 $\pm$   0.20 &     31.7 $\pm$      5.8  && 52162   &  1800. & $>$ 21.76\hfil & $<$     0.3 \hfil \\ 
751   &    10. &  15.72 $\pm$   0.19 &     34.3 $\pm$      5.9  &&   \\ 
\enddata
\tablenotetext{a}{time since trigger}
\tablenotetext{b}{flux in units of $10^{-16}$ erg cm$^{-2}$ s$^{-1}$ \AA$^{-1}$}
\end{deluxetable}

\begin{deluxetable}{r r r r r  }
\tablecolumns{5}
\tabletypesize{\footnotesize}
\tablewidth{0in}
\tablecaption{\label{Table4}UVOT White Light}
\tablehead{
\colhead{T(start)} &
\colhead{T(stop)} &
\colhead{Exposure} &
\colhead{V Mag\tablenotemark{a}} &
\colhead{V Flux\tablenotemark{b}} \\
\multispan3{\hfill (seconds)\hfill} \\
}
\startdata
    93167  & 111189 & 2378 &  21.35 $\pm$ 0.27 & 0.103 $\pm$  0.025  \\
   173859  & 329532 & 11750 & 22.82 $\pm$  0.43 &  0.027 $\pm$  0.010 \\
  329532  & 2599088 & 146267 & $>$ 24.11 &   $<$ 0.008    \\
\enddata
\tablenotetext{a}{equivalent V-band magnitude}
\tablenotetext{b}{equivalent V-band flux in units of $10^{-16}$ erg cm$^{-2}$ s$^{-1}$ \AA$^{-1}$}
\end{deluxetable}

The background subtracted 2--10 keV light curve in the 
time interval T+128~s -- T+1048~s (PD mode) is shown in Figure~3 (inset).
The X--ray afterglow of GRB 050525 is clearly fading. 
The early afterglow decay was first fitted with a single power-law model,
resulting in a best fit decay index $\alpha=-0.95\pm 0.03$, with  
$\chi_r^2 = 1.17$ (42 dof). Inspection of the residuals to the 
best fit model suggests that a flattening of the decay curve or a re-brightening of the source occurs 
at $\sim$300 seconds after the trigger.
A better fit is provided by a broken power law model with slopes 
$\alpha_1$,  $\alpha_2$ and a break at $t_b$. This model gave 
$\chi_r^2 = 0.98$ (40 dof), with best fit parameters 
$\alpha_1 = -1.23^{+0.03}_{-0.02}$, $\alpha_2 = -0.91$ and 
$t_b = 203~\mathrm{s}$. 

Again, however, the residuals suggest sytematic deviations from this model. We thus 
tried a broken power law with two temporal breaks. This model provided a very 
good fit to the data, with $\chi_r^2 = 0.72$ (38 dof) and is plotted in 
Figure~3 (inset) as a solid line. The best fit parameters are 
$\alpha_1 = -1.19$, $t_b^1 = 282~\mathrm{s}$, $\alpha_2 = -0.30$, $t_b^2 = 359~\mathrm{s}$, and
$\alpha_3 = -1.02$. 

Next, we fitted the X--ray data taken in PC mode at times more than 5000~s after the trigger. 
We first used a single power-law model,
obtaining a best fit decay index $\alpha=-1.51\pm 0.07$, with  
$\chi_r^2 = 1.40$ (12 dof). The poor fit is the result of a clear 
steepening of the light curve with time. We thus tried a broken power law 
model. The model provided a very
good fit with $\chi_r^2 = 0.97$ 
(10 dof) and best fit parameters $\alpha_1 = -1.16$, 
$\alpha_2 = -1.62$ and 
$t_b = 13177~\mathrm{s}$. 

Finally, we tried fitting the total light curve derived from the combined PD and PC mode data (see Figure~3).
We find that the power law fit to the pre-brightening PD mode data (T$<$280~s) extrapolates
well to the pre-break PC mode data. Moreover the decay index  before 280~s agrees well with
that of the PC mode data before the 13ks break. In contrast, if we extrapolate the post brightening
PD mode data to later times using the best fit slope, a significant excess is predicted compared
with the measured PC mode data. To join the post brightening PD mode data to the PC mode data 
requires a model with at least two temporal breaks, which are not constrained because of the 
intervening gap in X-ray coverage. We conclude that the brightening at about 280~s in the
PD mode data represents a flare in the X-ray flux, possibly similar to the sometimes much larger flares
that are seen at early times
in other bursts (Burrows et al 2005b; Piro et al. 2005), and that the flux returns to the pre-flare decay curve prior
to the start of our PC mode data. 

We thus fit the combined PD and PC mode data excluding PD data at times $t>$T+288~s (green points in Figure~3). 
A broken power law model 
provided a good fit (solid line of Figure~3), with with $\chi_r^2 = 0.50$ 
(25 dof) and best fit parameters $\alpha_1 = -1.20\pm 0.03$, 
$\alpha_2 = -1.62^{+0.11}_{-0.16}$ and 
$t_b = 13726^{+7469}_{-5123}~\mathrm{s}$. The break time is thus $\sim
3.8$ hours. 

The complete XRT data are recorded in Table~2.

\subsubsection{Spectral analysis}

The Photodiode spectra in the time intervals T+128~s - T+288~s
(``pre-flare'') and T+288s - T+1048s (``flare'') have been extracted in the 0.4--4.5 keV band. 
The spectra were binned to ensure a minimum of 20 counts per bin and were fitted using the XSPEC package 
(v.~11.3.2; Arnaud et al. 1996).

An absorbed power law model fits both spectra well.
For the ``pre-flare'' spectrum the best fit hydrogen-equivalent column density is 
$N_{\rm H} = 2.00\pm 0.25\times 10^{21}$~cm$^{-2}$ and the energy index is $\beta=-0.99\pm 0.09$ ($\chi_r2=0.94$, 176 dof). 
The best fit parameters for the ``flare'' spectrum are $N_{\rm H} = 2.07\pm 0.18\times 10^{21}$~cm$^{-2}$, 
$\beta=-0.96\pm 0.07$ ($\chi_r2=1.04$, 238 dof).

Freezing $N_{\rm H}$ to the Galactic value ($9.0\times 10^{20}$~cm$^{-2}$) and adding a hydrogen-equivalent 
column density at the redshift of the GRB
($z=0.606$)
we found $N_{\rm H}^z = 2.2\pm 0.5\times 10^{21}$~cm$^{-2}$, $\beta=-0.90\pm 0.08$ 
($\chi_r2=0.96$, 176 dof) for the ``pre-flare'' spectrum and $N_{\rm H}^z =  2.5\pm 0.4\times 10^{21}$~cm$^{-2}$, 
$\beta=-0.89\pm 0.06$ ($\chi_r2=1.01$, 238 dof) for the ``flare'' spectrum.

Events in PC mode from the time interval T+5859~s to T+41778~s 
were extracted from the same circular region used in the 
temporal analysis. A further selection on XRT event grades 0--4 (i.e.
single and double pixel events) was applied to the data. Again, the spectrum
was binned to ensure a minimum of 20 counts per bin. Energy channels below 
0.3 keV and above 10.0 keV were excluded.

An absorbed power law model fits  the PC data well
($\chi_r^2=0.70$, 21 dof), with 
$N_{\rm H} = 1.5^{+0.8}_{-0.7}\times 10^{21}$~cm$^{-2}$ and 
$\beta=-1.10^{+0.27}_{-0.24}$.

No evidence is found for evolution of the X-ray spectral shape with
time (cf. Chincarini et al. 2005).

\begin{figure}
\includegraphics[angle=-90,scale=0.5]{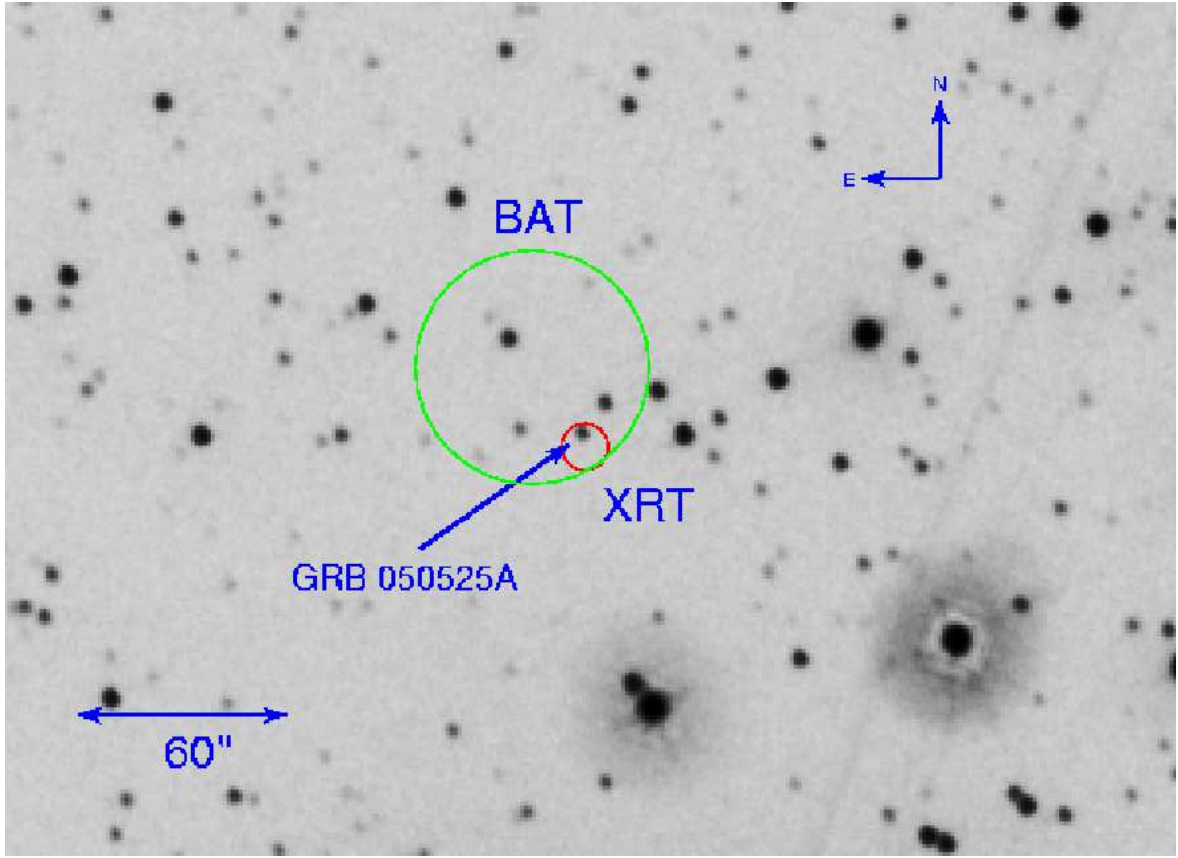}
\caption{This Figure shows the field containing GRB 050525A, marking the location of the optical afterglow (the
object picked out by the arrow).  The BAT 
error circle 
is based on the refined ground-based analysis (Cummings et al. 2005), and this 
has a 
radius of 0.5 arcminutes.  The XRT error circle has a radius of 6 
arcseconds and the central position has been revised compared to that reported in Band et al. (2005) after ground analysis.
}
\end{figure}

\begin{figure}
\plotone{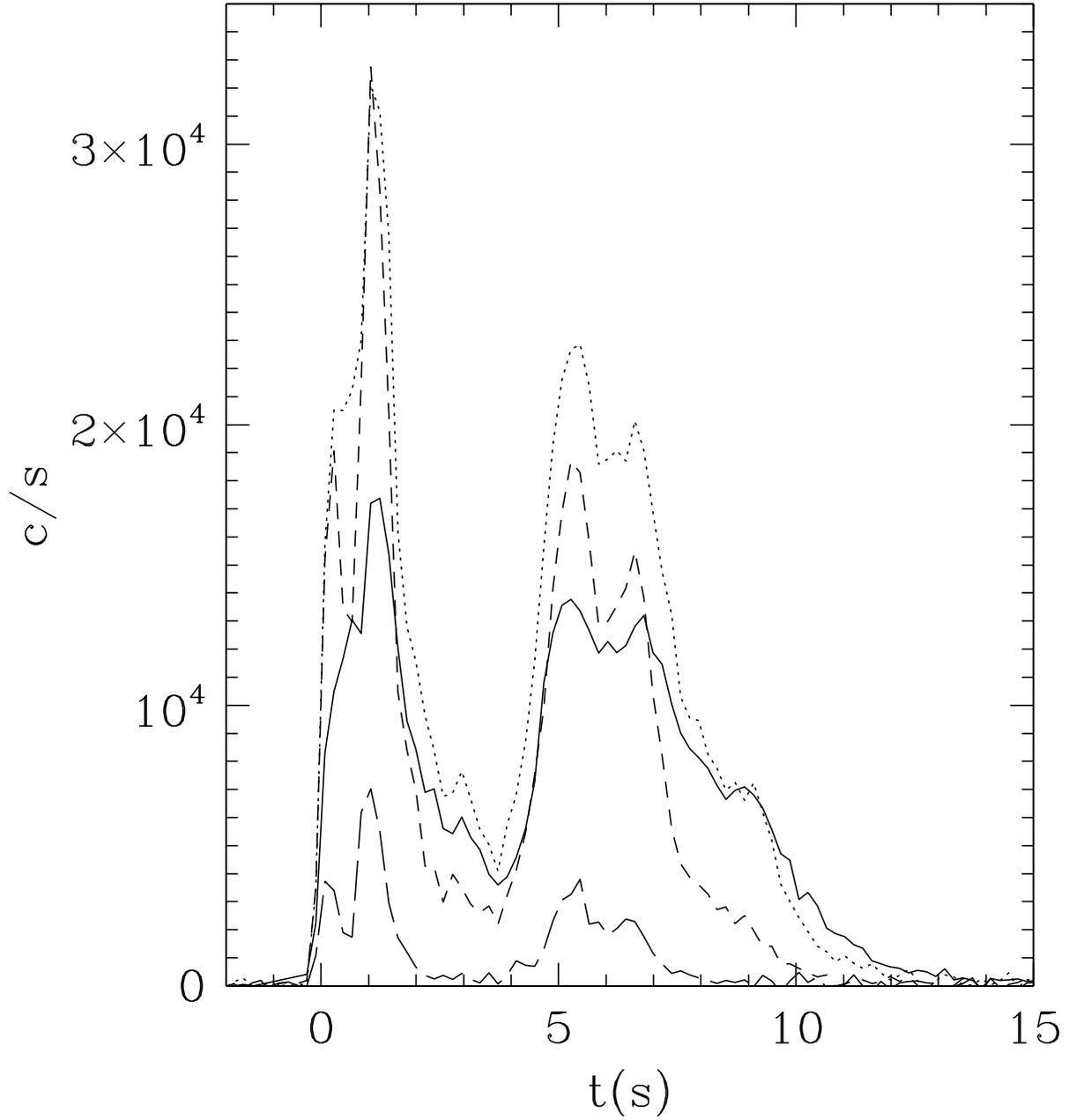}
\caption{
BAT lightcurve of GRB050525a in the 15-25 keV (solid), 25-50 keV (dots), 50-100 keV (short dashes), 
and 100-350 keV (long dashes) bands. 
}
\end{figure}

\begin{figure}
\epsscale{0.8}
\plotone{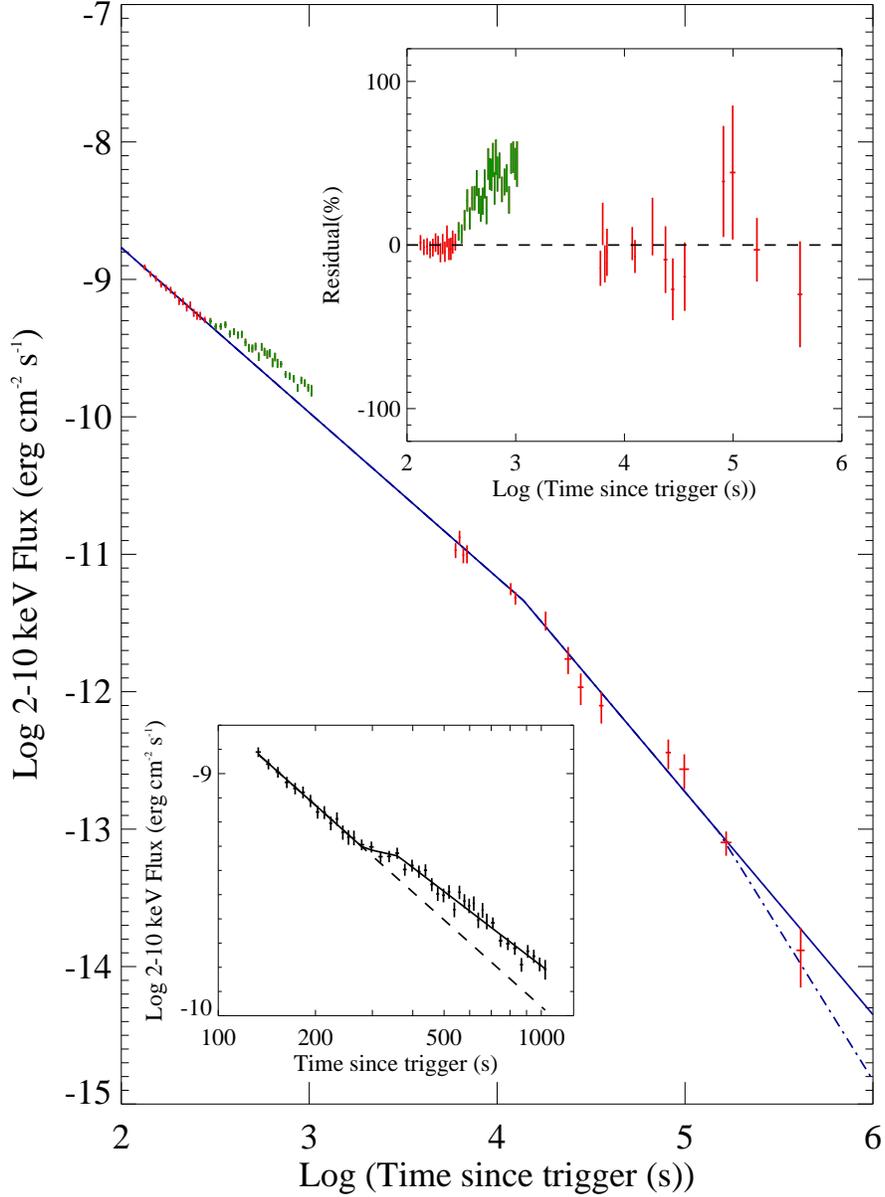}
\caption{XRT decay light curve of GRB050525a including both Photodiode Mode (T$<$2000~s) and Photon Counting Mode 
(T$>$2000~s) data. The solid line is a broken power law fit to the combined data excluding those Photodiode Mode points 
colored green (see text). The dash-dot line is shown for illustration and has a slope of $\alpha=-2.2$, which is the value expected from simple 
modelling of a jet break (see text).  The {\sl lower inset} shows the data taken in photodiode mode only,
during the first $\sim$1000~s after the BAT trigger. The solid line is a fit to the data with a power law model
that includes two temporal breaks to different decay rates. The dashed line is an extrapolation of a 
simple power law fit (single slope) to the first segment of data prior to about 300~s.
The {\sl upper inset} shows the residuals with respect to the two power-law model fit to all the data, expressed 
as a percentage 
of the predicted model 
flux.}
\end{figure}


\begin{figure}
\epsscale{1.0}
\plotone{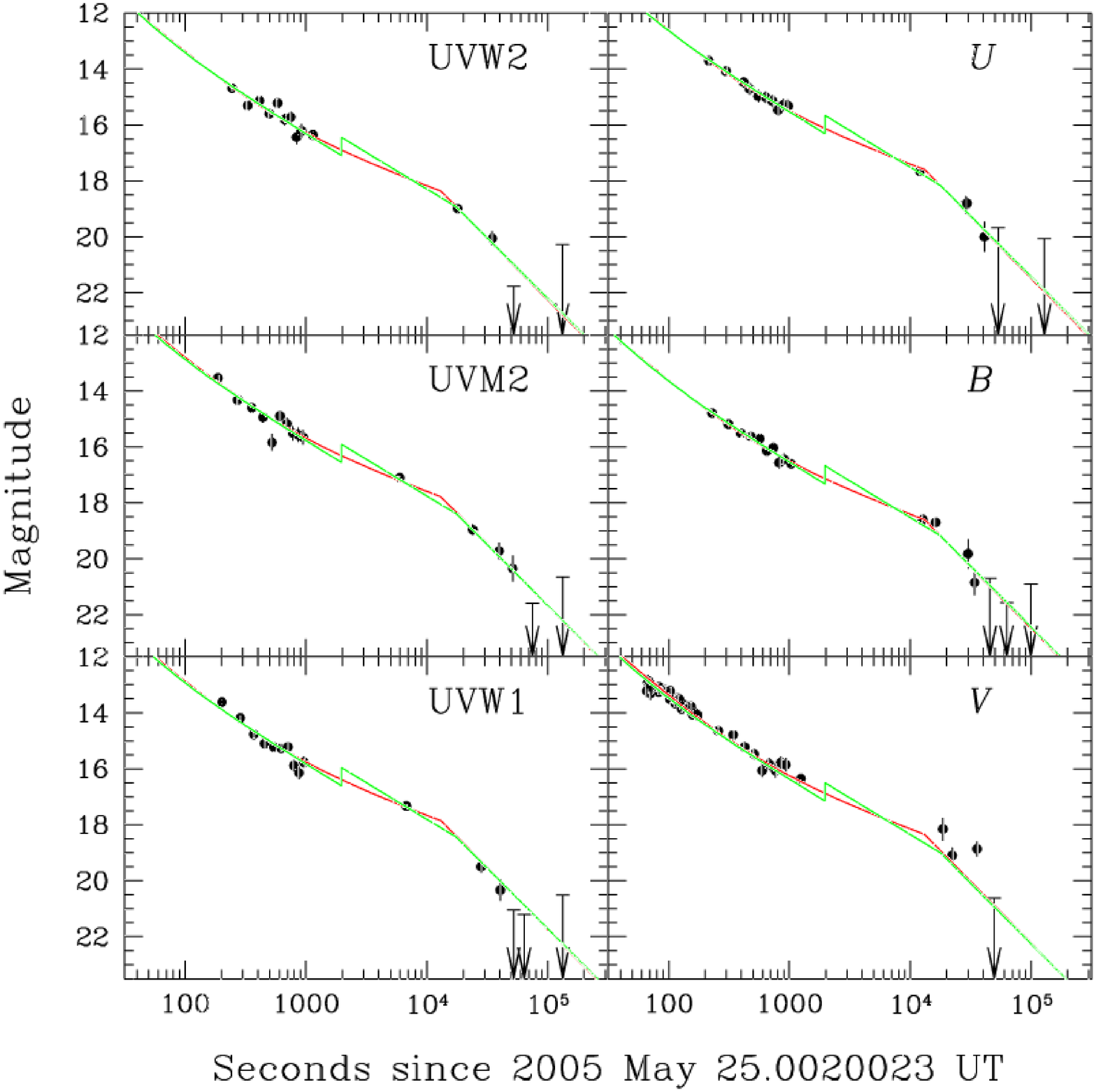}
\caption{    
    This Figure shows the photometry (solid circles) in the six UVOT
broadband filters.  Arrows indicate 3$\sigma$ upper limits to the
photometry.  A model light curve (red line) with $\alpha_1 = -1.56$,
$\alpha_2 = -0.62$, $\alpha_j = -1.76$, $t_t = 432$ s (7.2 min), and
$t_j = 13\,133$ s (0.152 days) is shown.  The green line indicates a
rebrightening model with $\alpha_1 = -2.14$, $\alpha_2 = -1.04$,
$\alpha_j = -1.73$, $t_t = 39$ s, and $t_j = 17\,747$ s (0.205 days).
The flux has been normalized for each filter.
}
\end{figure}

\begin{figure}
\epsscale{1.0}
\plotone{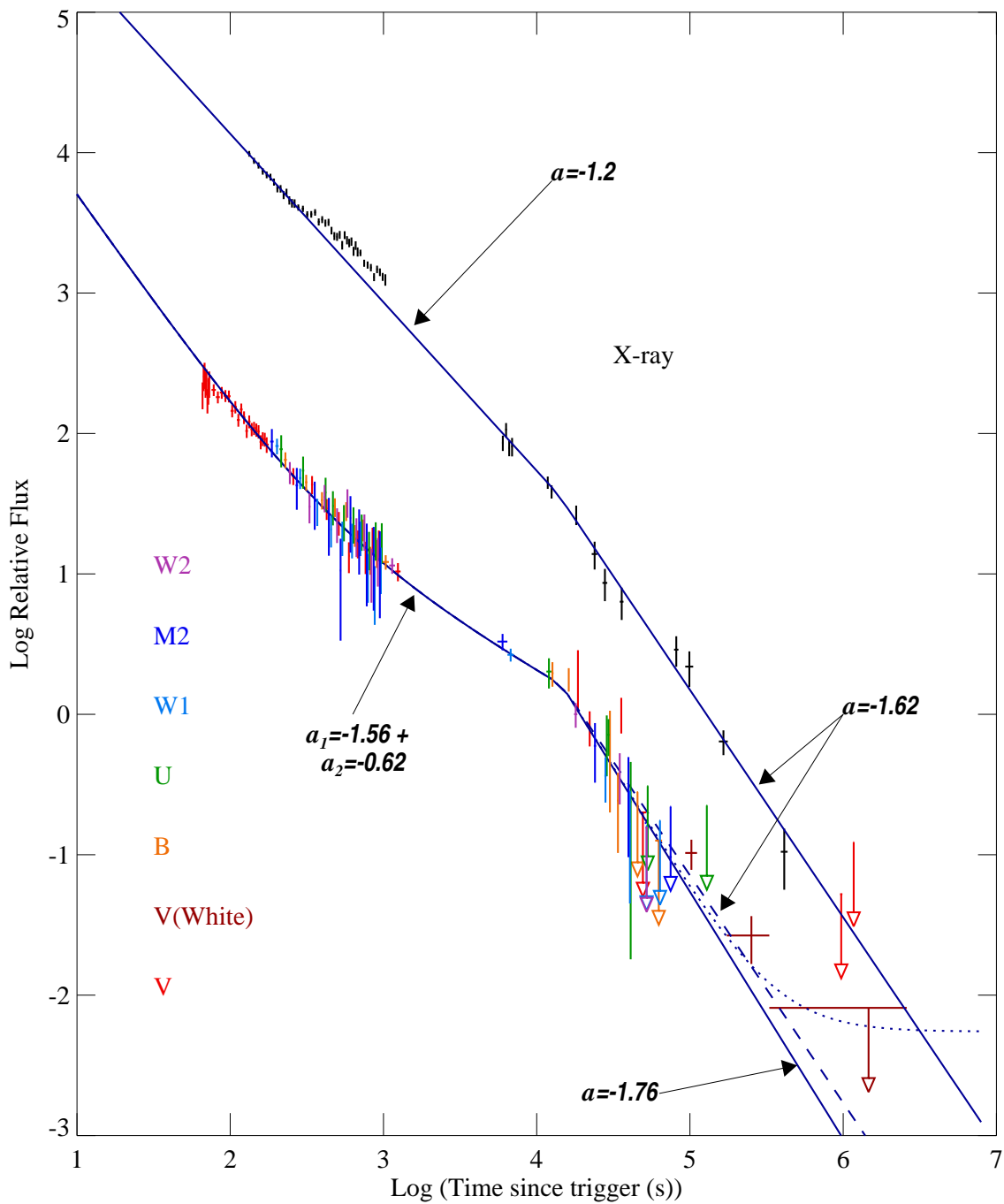}
\caption{    Comparison of flux decay in the X-ray and UVOT bands. The UVOT data have been normalised 
in the interval up to T+1000s, and the data taken through different filters are distinguished by color. 
The relative normalisation of the X-ray and optical/UV data is arbitrary. The best fit
broken power law model is plotted through the X-ray data. The best fit double power law with break 
is plotted through the UVOT data (see text). The dashed line has the same post break slope as the X-ray data. The dotted line
is the best fit model with a constant flux added corresponding to the value measured by 
Soderberg et al. (2005) using HST/ACS  (see text).
}
\end{figure}

\begin{figure}
\plotone{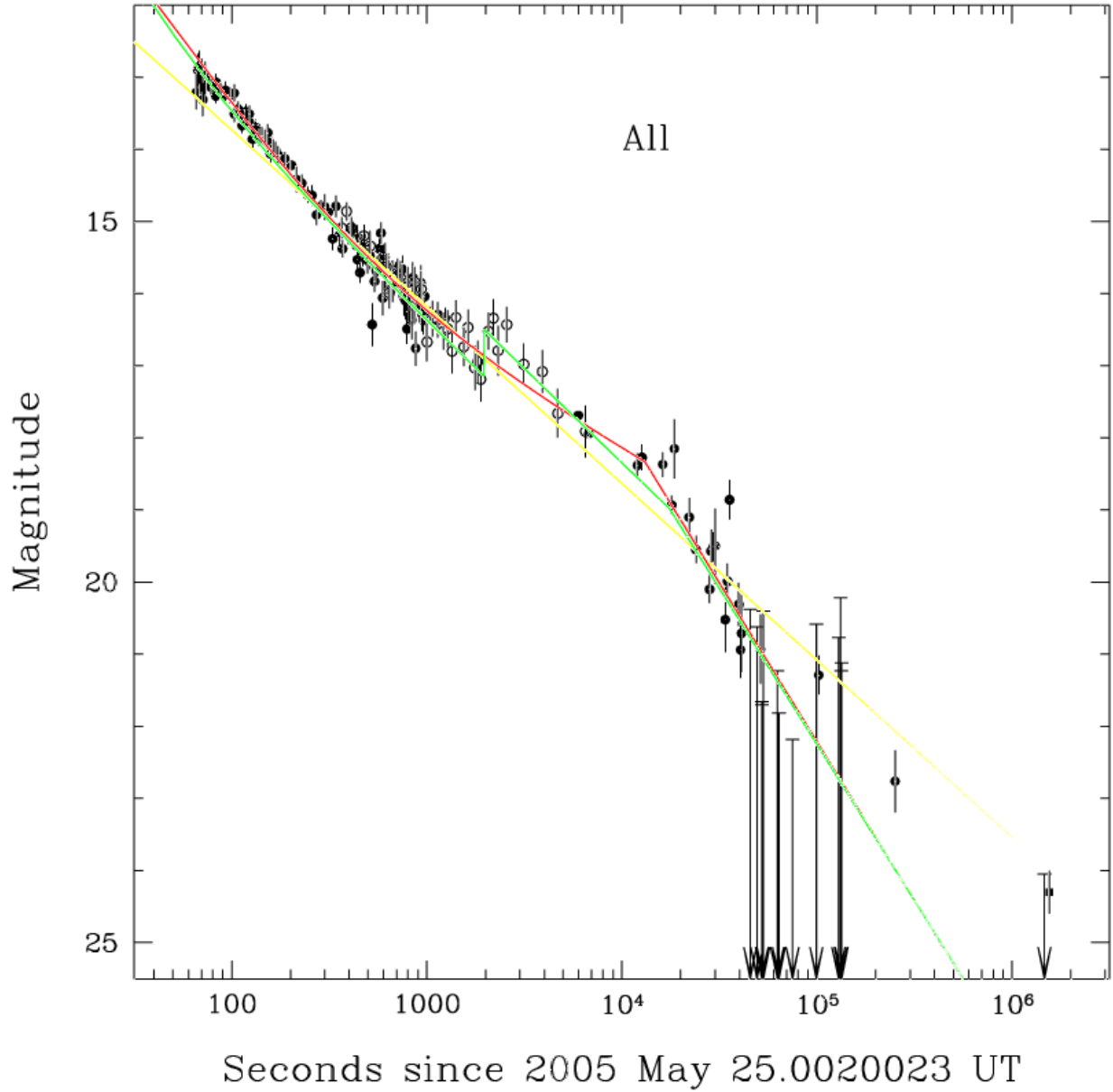}
\caption{        This Figure shows the UVOT photometry (solid circles) normalized
so that the early time flux is the same in each filter.  The red and
green lines are the models from Figure~4.  The yellow line is the
best-fitting single power law.  The open circles are the
Klotz et al. (2005) data scaled to the same flux scale.  The solid
square is the {\sl HST\/}/ACS F625W data of Soderberg et al. (2005).  Note that
the {\sl HST\/}/ACS data point is approximately three mag brighter
than the predicted afterglow magnitude.
}
\end{figure}

\begin{figure}
\epsscale{0.8}
\plotone{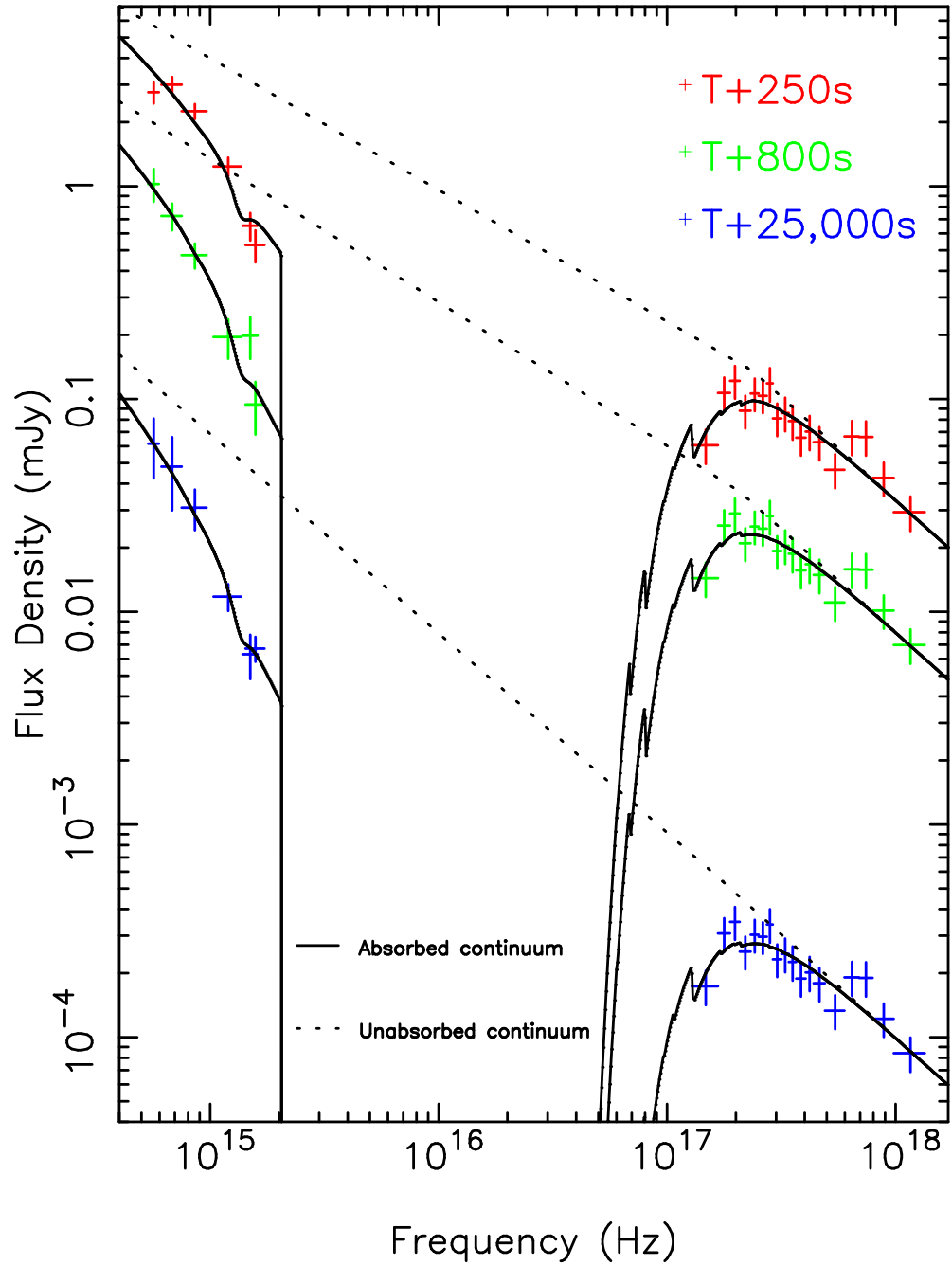}
\caption{    UVOT and XRT data interpolated to the epochs T+250s, T+800s and
T+25,000s, and best spectral fit models (solid lines). The dotted lines
represent the intrinsic continuum of the source, before extinction and absorption 
from gas and dust in both the Milky Way and the host galaxy.
}
\end{figure}



\subsection{UVOT data}
The Swift UVOT began taking data in the V filter $\sim 65$ seconds after the 
burst trigger, immediately after the GRB came into its field of view. After about ten seconds
the spacecraft attitude had settled and UVOT began a 100s `finding chart' exposure in V. 
Thereafter the instrument cycled through each of six color 
filters, V, B and U 
together with filters defining three ultraviolet passbands, UVW1, 
UVM2 \& UVW2 with central wavelengths of 260nm, 220nm \& 193nm respectively. 
The exposure duration per filter was initially 10 seconds, 
subsequently increasing to 100s and then 900s at 
predetermined times after the trigger. Data were taken in both 
`event mode', in which the time and detector position of each 
individual photon is recorded, and in `image mode', where the image is accumulated
onboard, discarding the photon timing information within an exposure to reduce telemetry volume.
At some phases of the observation both modes were operated simultaneously through different sized spatial windows 
that were
selected in combination with the spatial binning of the image mode data 
so as to match the overall data volume with the available telemetry rate.
Event mode, and larger spatial windows, are emphasised at the beginning of the observing sequence, when the afterglow brightness
is changing rapidly and its position is less well known. Later in the sequence only image mode data are taken.
The intrinsic spatial 
pixel size of the detector is approximately 0.5 arc seconds on the sky, but some image mode data were taken 
with the data binned 2x2 
to give 1 arc second pixels.

Examination of the UVOT finding chart exposure reveals a bright new fading source within the XRT positional error 
circle of the burst at RA = 18h 32m 32.62s Dec = +26d 20' 21.6'' (J2000) with an 
estimated uncertainty of 0.2''. This close to the ROTSE position given by Rykoff et al. (2005).

In Figure~4 we show the light curves in each of the UVOT bands as a function of time. These data
were derived from measurements made in a 6 arcsec aperture, with background determined from a nearby source-free region. 
Since the UVOT detector counts individual photons,
it is subject to coincidence loss (pile-up), which is noticeable for count rates above 10 per second. We have 
applied the standard coicidence loss correction to the data, derived from ground calibrations (Breeveld et al. 2005). The data have
also been corrected for Galactic extinction, adopting a value of E(B-V)=0.095 for this field 
(Schlegel, Finkbeiner \& Davis 1998). We used the data of Schlegel et al. (1998) and Pei et al. (1992) to
derive extinction values for each filter. Expressed as magnitudes these are  
0.688 (UVW2), 0.926 (UVM2), 0.740 (UVW1), 0.518 (U), 0.412 (B) \& 0.316 (V). The data are tabulated in Table~3, 
separated by filter. The detector count rates have been converted to magnitudes and fluxes based on in-orbit
calibration data. Systematic uncertainties between the filters are estimated to be better than 0.15 magnitudes.

After the source
had faded below the threshold of detectability through the UVOT color filters, we made observations in unfiltered (white) 
light. The white light data provided detections of the afterglow beyond $10^{5}$ s. We place the white light
data on the V magnitude and flux scale using a color dependant correction that is derived by relating the measured
ratio of V to white counts of field stars and photometric standards to their apparent B-V
color (i.e. B-V color not corrected for
reddening). The white light data are contained in Table~4.

\subsubsection{Temporal Analysis}

We fit a single power law
to the light curve for each UVOT broadband filter individually, initially
using only the data at $t \le 1000$ s after the BAT
trigger where the sampling for each individual filter is relatively dense. 
Each filter can be reasonably fit to a common slope of $\alpha =
-1.14 \pm 0.01$.  We conclude that there is no evidence for a gross change in
the optical/UV spectrum associated with the temporal decay.

To maximise the light curve sampling, we combine the data from
all the UVOT filters, multiplying the individual curves by a factor that normalises them to 
a common flux scale.
The results are shown in Figure~5.
We also include for comparison the X-ray decay curve of Figure~3, which has been multiplied by a constant factor in
Figure~5 for display purposes. 
It can be seen that the X-ray and optical/UV signals initially fade at a similar rate, but
that the optical/UV flux flattens compared to the X-ray curve in the interval 1000-10000s (i.e. the ratio of
optical to X-ray flux increases with time). After about 10000s the optical curve steepens in a similar 
way to the X-ray curve.

The complex nature of the optical/UV light curve is confirmed by the fact that a
single power law does not provide a good fit to the combined UVOT data over the entire
time range.  The
chi-square for the fit is 1038 for 115 degrees of freedom. We instead tried various combinations
of power-laws
and found that
the best
fit was obtained when we fit three power law components
simultaneously.  The model consists of the sum of two power laws 
(Holland et al. 2004) that breaks to a third, single power law at late time.  This
is physically consistent with the flux being dominated by a reverse
shock at very early time, a forward shock at intermediate time, and a
jet break at late time.  The best fitting model had an early-time
power law decay of $\alpha_1 = -1.56^{+0.07}_{-0.08}$ transitioning to
a power law with a slope of $\alpha_2 = -0.62^{+0.05}_{-0.04}$ at $t_t
= 432^{+173}_{-86}$ s.  This transition time is the time at which both
power law components contribute equally to the observed flux from the
afterglow.  An achromatic break occurs at $t_j =
13\,133^{+691}_{-691}$ s after the BAT trigger.  We interpret this as
a jet break.  The decay index after the jet break is $\alpha_j =
-1.76^{+0.08}_{-0.08}$.  The time of the jet break, and the post-break
index, are consistent with the values from the $X$-ray data.  The
chi-square value for this model is 555 for 111 degrees of freedom,
the high value reflecting significant short-term scatter of individual points about the model.
This model is compared to the individual filter
data in Figure~4.

     Klotz et al. (2005) find that the $R$-band light curve of the 
GRB~050525A afterglow underwent a sudden rebrightening of 0.65 mag
at 1968 s after the BAT trigger. This was at a time when GRB050525a was not visible to Swift. 
We show the Klotz et al. data in Figure~6, normalised to
the UVOT curve. 
If we add this rebrightening to our
model the best fit has $\alpha_1 = -2.14^{+0.25}_{-0.26}$,
$\alpha_2 = -1.04^{+0.02}_{-0.02}$, $\alpha_j =
-1.73^{+0.08}_{-0.11}$, $t_t = 39^{+6}_{-6}$ s, and $t_j =
17\,747^{+1523}_{-1172}$ s with a chi-square of 597 for 111 degrees of
freedom.  Although the chi-square value is formally larger with a
rebrightening, an examination of Figures~4 \& 6 suggests that both models
provide comparable fits.

Soderberg et al. (2005) have reported a late-time measurement of the GRB050525a afterglow brightness using HST/ACS.
This is illustrated in Figure~6. The HST measurement is approximately 3 magnitudes brighter than predicted
at this time by an extrapolation of our best fit afterglow model. If the HST measurement represents a
constant background flux, for example due to the host galaxy or slowly varying supernova emission, then
we estimate that this would contribute about 25\% of the flux that we measure in our UVOT white 
light detection at $\sim 2.5 \times 10^5$s. The effect of such a constant additional flux on the best fit 
model is illustrated in Figure~5 as the dotted line.

\subsubsection{Multiwavelength Spectral Analysis}

Combining data across the Swift instruments provides a powerful set of diagnostcs for prescribing 
the instantaneous spectral properties of a source and its temporal evolution.  In order to fit an 
instantaneous spectral model to the UVOT data, it is necessary to interpolate count rates through each 
filter to a common epoch. We build a broad-band spectrum of the source at T+250s using the appropriate 
decay index for each filter, and the nearest exposure to T+250s as a normalization reference. The XRT 
spectrum described in section 2.2.2 was also renormalized to T+250s according to the early decay slope from 
Figure~3.
As prior assumptions within the spectral model we take the XRT slope and 2--10 keV flux determined in section 2.2.2, the Galactic 
Hydrogen column density in the direction of the source of $9 \times 10^{20}$ cm$^{-2}$ (Dickey \& Lockman
1990) and the Galactic extinction in the same direction characterized by Pei (1992) with E(B-V) = 0.095 
(Schlegel et al 1988) and $R_V = 3.1$. Applying this model directly to the combined UVOT and XRT spectrum 
at T+250s yields a poor fit with $\chi^2 = 5.3 \times 10^{4}$ for 24 dof. This particular model systematically 
over-predicts the optical and UV flux. We attempt to correct this by first adding dust extinction and gas 
absorption from the host galaxy at a fixed redshift of $z = 0.606$. The dust is initially assumed to have a 
content identical to the SMC with $R_V$ = 2.93 (Pei 1992). This does not provide an acceptable fit, with 
$\chi^2$ = 272 for 22 dof, where
$N_{\rm H}(host) = (1.8 \pm 0.8) \times 10^{21}$ cm$^{-2}$ and E(B-V)(host) = $0.43 \pm 0.03$.  
Adopting host dust characteristics identical to the Galaxy and LMC populations (Pei 1992) 
does not improve the fit significantly, yielding $\chi^2$ = 208 and 242, respectively for 22 dof. 
An acceptable fit is, however,  obtained if we add a spectral break to the model at 1 keV. Details of the fits to the
model with a spectral break are provided in Table~5. The table includes results obtained using 
SMC, LMC and Galaxy dust distribitions. The SMC prescription 
is preferred statistically over the other two.

\begin{deluxetable}{l l c c c}
\tablecolumns{5}
\tabletypesize{\footnotesize}
\tablewidth{0in}
\tablecaption{SED at T+250s}
\tablehead{
 & & 
\colhead{SMC} &
\colhead{LMC} &
\colhead{Galaxy} \\ 
}
\startdata
$N_{\rm H}$(host)& (cm$^{-2}$) & $1.2 \pm 0.8 \times 10^{21}$ &  $1.2 \pm 0.8 \times 10^{21}$ & $1.2 \pm 0.8 \times 10^{21}$ \\
$\beta _1$	& &	 $-0.60 \pm 0.04$	&   $-0.63 \pm 0.03$	 &  $-0.62 \pm 0.04$ \\
$\beta _2$\tablenotemark{*}	& &	 $-0.97$ 	 & 	  $-0.97$ 	&   	   $-0.97$ \\
E$_{\rm break}$  & (keV)	& $1.0 \pm 0.2$ &	  $1.0 \pm 0.2$  & $ 1.0 +/- 0.2$ \\
E(B-V)(host) & &	 $0.04 \pm 0.02$ &	  $0.06 \pm 0.03$ &	   $0.06 \pm 0.04$ \\
$\chi^2$(dof) & &	 27(20)	&  	  31(20)  & 	   41(20) \\
\enddata
\tablenotetext{*}{Fixed parameter}
\tablecomments{Best fit spectral parameters for the Spectral Energy Distribution (SED) of GRB 050525a at
T+250s for three dust populations assumed for the host galaxy. The
model continuum is a broken powerlaw ($\beta _1$ where E $<$ E$_{\rm break}$ and $\beta _2$ where 
E $>$ E$_{\rm break}$). The slope in the X-ray band, $\beta _2$ = 0.97, and the 2--10 keV flux are fixed at the XRT determined 
values. Galactic values for gas absorption and dust extinction are used, while the 
gas and dust properties of the host galaxy (N$_{\rm H}$(host) and E(B-V)(host)) are allowed to float.
}

\end{deluxetable}

We then perform spectral fits using the broken powerlaw model with SMC-like dust in the host galaxy for 
two other epochs, T+800s and
T+25000s. Data and best-fit models are plotted in Figure~7 while
best-fit parameters are listed in Table~6.  Other than temporal decay, we 
find no evidence for spectral evolution between 200s and 800s after the burst; the spectral 
break has not moved within errors. However at
T+25,000s the spectrum no longer requires a break between the XRT and
UVOT energy bands. The UVOT spectral index is identical to the XRT index, within 
measurement uncertainties. There is no evidence for evolution in the host's gas or dust enviroments over these epochs.

\begin{deluxetable}{l l c c c}
\tablecolumns{5}
\tabletypesize{\footnotesize}
\tablewidth{0in}
\tablecaption{The SMC-dust model from Table 5 applied to the GRB 050525a SED at three epochs.}
\tablehead{
 & & 
\colhead{T+250s} &
\colhead{T+800s} &
\colhead{T+25,000s} \\ 
}
\startdata
N$_{\rm H}$(host)& (cm$^{-2}$) & $1.2 \pm 0.8 \times 10^{21}$ &  $1.5 \pm 0.9 \times 10^{21}$ & $1.8 \pm 1.0 \times 10^{21} $\\
$\beta _1$	& &	 $-0.60 \pm 0.04$	&   $-0.67 \pm 0.06$	 &  $-0.94 \pm 0.10$ \\
$\beta _2$\tablenotemark{*}	& &	 $-0.97$ 	 & 	  $-0.97$ 	&   	   $-0.97$ \\
E$_{\rm break}$  & (keV)	& $1.0 \pm 0.2$ &	  $0.9 \pm 0.2$  & -- \\
E(B-V)(host) & &	 $0.04 \pm 0.02$ &	  $0.09 \pm 0.04$ &	   $0.08 \pm 0.05$ \\
$\chi^2$(dof) & &	 27(20)	&  	  18(20)  & 	   15(20) \\
\enddata
\tablenotetext{*}{Fixed parameter}
\end{deluxetable}

\section{Discussion}

GRB050525a is the second most fluent GRB to have been observed by Swift to date, and is the first bright low redshift
burst to have been observed since all three Swift instruments have been operational. The hard X-ray/Gamma-ray 
temporal charateristics and spectrum
of this bright burst have been well characterised by the BAT. Added to this, the XRT and UVOT instruments provide well
sampled temporal multiwavelength decay light curves and spectra starting about 1 minute after the burst trigger and 
extending to several days after the burst, making this one of the best covered GRBs thus far. 
The ratio of the gamma-ray fluence to the X-ray flux is typical of the
general population of long bursts observed in the pre-Swift era (cf. Roming et al. 2005b).

In an initial period of about 4 hours after the 
burst, the decline of the X-ray afterglow flux of GRB050525a can be well 
represented by a simple power law
with an index $-1.2$, apart from a probable flare which occurs after about 300s. 
The optical temporal signature is more complex, however, and exhibits a significant flattening
over that same period. It can be reproduced by the combination of two power laws (or two power laws combined with 
a re-brightening episode), one declining more rapidly than the X-ray flux, and the other more slowly. This can
be interpreted as the combintion of a steep reverse shock component, combined with the flatter optical apparition of the 
forward shock seen in X-rays. The X-ray and optical/UV multiwavelength spectral fits suggest that
the cooling frequency of the electrons in the forward shock has moved through the optical band by 25000s 
after the burst.
There is good evidence for a break in the 
time evolution of both the X-ray and optical/UV data at 13000-14000s ($\sim$0.15 days) after the trigger. The break is 
consistent with being achromatic i.e. the break time is, within errors, consistent between the X-ray and optical 
bands, and the post-break slope is also consistent within the errors (cf. sections 2.2.1 and 2.3.1). Its 
achromatic nature 
suggests that this is the so-called jet break, which occurs when the 
beaming angle of the decelerating relativistic flow approaches the colimation angle of the jet (Sari, Piran \& Halpern 1999). 

Comparing the temporal and spectral indices in both the UVOT band and the XRT band with the theoretical models 
(e.g. Table 1 in Zhang \& M\'esz\'aros 2004) as well as the ordering of various temporal segments on the X-ray 
and optical lightcurves (Sari et al. 1998; Chevalier \& Li 2000), we find that the global data are not consistent 
with a model that invokes a massive stellar wind.  This is also supported by the fact that the spectral break 
energy $E_{break}$ in our joint XRT-UVOT spectrum (which is consistent with a cooling break) evolves downwards 
with time, since $\nu_c$ increases with time in the wind model. The UVOT-XRT data before $\sim 400$s could be 
accommodated with a slow-cooling wind model in which the UVOT band is between $\nu_m$ and $\nu_c$ 
(with the temporal index $(1-3p)/4$) while the X-ray band is above $\nu_c$ (with the temporal index $(2-3p)/4$). 
However, the $\nu_m < \nu <\nu_c$ segment is expected to happen at a later time, and it is typically the 
last segment in the lightcurve so that it is very unlikely that there would be a switch back to a shallower decay in the 
wind model (Fig.1 of Chevalier \& Li 2000). The early steep-to-shallow transition observed in the UVOT 
lightcurve naturally rules out the wind interpretation, and is consistent with a reverse-forward shock 
model as discussed below. Another possibility is that the fireball may be initially in a wind medium, but 
later runs into an ISM after it crosses a wind-termination shock. An additional motivation for this 
interpretation is the optical-rebrightening at $\sim 2000$s revealed by the TAROT data (Klotz et al. 2005), 
which could be presumably the signature of the termination shock. Although the possibility that the 
rebrightening is due to a density enhancement in the medium surrounding the burst is promising, the above scenario lacks direct proof since 
the wind signature, if it is one, terminates at a much earlier time (e.g. $\sim
400$s) than the epoch of the optical rebrightening bump.

Instead, the XRT and UVOT data before the presumed jet
break at 0.15 d are in good agreement with the standard fireball model for a constant ISM density 
(e.g. Sari, Piran \& Narayan 1998). According to this model, the spectral indices ($\beta$) above 
and below the synchrotron cooling frequency $\nu_c$ are $-p/2$ and $-(p-1)/2$, respectively; while 
the temporal indices ($\alpha$) for a band which is above and below $\nu_c$ are $(2-3p)/4$ and $(3-3p)/4$, 
respectively. Interpreting $E_{break}$ as the cooling break, and taking $p=2.2$, one expects $\alpha_{X}=-1.15$, 
$\alpha_{O} =-0.9$, $\beta_X=-1.1$, and $\beta_O=-0.6$. All these are in general agreement with the data, particularly when 
the optical re-brightening at $\sim 2000$s (Klotz et al. 2005) is taken into consideration when fitting model parameters. 
At $T+25,000$s, the energy spectrum of the burst is 
consistent with a single power law spectrum extending between the XRT and UVOT bands, with index $\beta \sim -0.95$. 
This is consistent with the fact that the cooling frequency has already crossed the UVOT band at this epoch. 
The  decay slope expected in the UVOT band at this time should be $\sim -1.15$ in the standard model. The observed 
decay slope, however, is steeper ($\sim -1.6$). This is due to the jet 
break around $\sim T+14000$s. According to the standard afterglow model, $\nu_c$ evolves with time 
as $\propto t^{-1/2}$. The data, on the other hand, requires a slightly faster evolution of $E_{break}$. 
Enhanced cooling is thus needed, possibly with an evolving $\epsilon_B$, the magnetic equipartition 
parameter in the shock. An evolving $\epsilon_B$ was also invoked to interpret data GRB 050128 (Campana et al. 2005).

A temporal break is identified in the X-ray lightcurve at around 14000s (0.15d). The UVOT data show a break around 
the same time. Since there is no apparent spectral change aross the break, and since the break time 
is ``achromatic", this points to a jet break, corresponding to the time when the fireball Lorentz 
factor $\gamma \sim 1/\theta_j$ (Rhoads 1999; Sari, Piran \& Halpern 1999). The post-break temporal 
index, however, is shallow ($\sim -1.6$) compared with the model prediction (which is $\sim -p = -2.2$). 
Such a shallow jet break has also been seen previously in some other bursts (e.g. GRB 990123, Kulkarni et al. 1999). 
It might be shallow because the sideways expansion effect is not significant (Panaitescu \& M\'esz\'aros 1999). 
Another possibility is that the jet break turn-over time may be finite (Panaitescu \& M\'esz\'aros 1999), so 
that the post-break asymptotic regime has not yet been reached. If this is the case, then the effective 
jet break time could be later than we have fitted. For example, while the increased number of free parameters are 
not justified by the quality of our data, we could force a 3-power-law fit to our X-ray data with a second break to
a slope of $\alpha = -2.2$ at about T$= 1-1.5 \times 10^5$s, yielding an `effective' single break 
time of about 50,000-60,000s. This is illustrated in Figure~3, where the dash-dot line shows a break to a slope 
of $\alpha = -2.2$ at T$= 1.5 \times 10^5$s. The final two XRT points in isolation 
are equally consistent with a slope of $\alpha=-1.6$ and $\alpha=-2.2$. 
It is also still possible that the observed 
temporal break is due to reasons other than a jet break (e.g. Gendre \& Bo\"er 2005), but given that this is the first 
time one sees an achromatic break in both the X-ray and the optical band, we tentatively conclude 
that the current data support a jet break interpretation.

If the break is indeed attributed to a jet, one can derive
a jet angle (e.g., Sari et al. 1999)
\begin{equation}
\theta_j=0.12 \left(\frac{t_j}{1+z}\right)^{3/8}
    \left(\frac{n_0 \eta_\gamma} {E_{\rm iso,53}} \right)^{1/8} =
    5.6 \times 10^{-2} \left(\frac{n_0 \eta_\gamma}{0.6} \right)^{1/8}
    \left(\frac{t_j}{\hbox{0.15 day}}\right)^{3/8}~, \end{equation} %
where $t_j$ is expressed in days, $n_0$ is the ambient density, $\eta_\gamma$ is the ejecta-to-gamma-ray efficiency, 
and $E_{\rm iso,53}$ is the isotropic energy in units of $10^{53}$~erg.  In our calculations we use 
$n_0 = 3$~cm$^{-3}$ and $\eta_\gamma = 0.2$ (Ghirlanda, Ghisellini \& Lazzati 2004).  Therefore 
$\theta_j\sim 3.2^\circ$ if we use a jet break time of $t_j\sim 0.15$~day, or $\theta_j\sim 5.4^\circ$ 
if we adopt a more conservative jet break time of $t_j\sim 0.6$~day corresponding to a finite roll-over 
as discussed above; these two values bracket the possible range of $t_j$. This angle could be interpreted 
as the physical opening angle of a jet if the jet has a uniform distribution of energy, or as the observer's 
viewing angle with respect to the jet axis in a structured jet model (Rossi et al. 2002; Zhang \& M\'esz\'aros 2002; 
Kumar \& Granot 2003). Within the uniform jet model, we estimate the actual gamma-ray energy emitted to be 
$E_\gamma = E_{iso,\gamma} (1-\cos\theta_j) = 3.6 \times 10^{49}$ ergs for $t_j\sim 0.15$~day or 
$E_\gamma = 1.0 \times 10^{50}$ ergs for $t_j\sim 0.6$~day.


We can combine the information on the jet with the
measurements of the redshift, $E_p$, $E_{iso,\gamma}$, $t_j$ and the lag between the 
high energy and low energy gamma-ray light curves (section 2.1) to test whether this 
burst is consistent with various proposed empirical relations. Four relations involve 
$E_{\rm pt}$, the peak energy in the burst frame for the entire burst:  the Amati 
relation---$E_{\rm pt,Amati} \propto E_{\rm iso}^{0.5}$ (Amati et al. 2002); the 
Ghirlanda relation---$E_{\rm pt,Ghirlanda} \propto E_\gamma^{0.7}$ (Ghirlanda et al. 2004); 
the Yonetoku relation---$E_{\rm pt,Yonetoku} \propto L_{\rm iso}^{0.5}$ (Yonetoku et al. 2004); 
and the Liang-Zhang relation---$E_{\rm pt,LZ} \propto E_{rm iso}^{0.52} 
t_j^{0.64}$.
(Liang \& Zhang 2005).  For
GRB~050525a's values of $E_{\rm iso}$, $E_\gamma$, $t_j$ and $L_{\rm iso}$, we calculate 
$E_{\rm pt,Amati}$=144~keV, $E_{\rm pt,Ghirlanda}$=46~keV (if $t_j=0.15$~day, 96~keV if $t_j=0.6$~day), 
$E_{\rm pt,Yonetoku}$=186~keV, and $E_{\rm pt,LZ}$=50~keV (if $t_j=0.15$~day, 122~keV if $t_j=0.6$~day), 
which should be compared to the observed value of $E_{\rm pt}=126.6\pm5.5$~keV.  The set of bursts used to 
calibrate these relations is small enough that each additional burst with a redshift and a spectrum will 
change the relations quantitatively; in addition, there is significant scatter around these relations 
(G.~Ghirlanda, 2005, personal communication).

According to the lag-luminosity relation, shorter lags between the
emission in two gamma-ray bands are correlated with larger
$L_{\rm iso}$ (Norris, Marani \& Bonnell 2000).  Using the
methodology in Band, Norris \& Bonnell (2004), we calculate
a redshift $\hat z = 0.69 \pm 0.02$ for GRB050525a from the observed
values of the peak flux, lag and spectrum at the peak of
the lightcurve (Norris et al. 2005); the uncertainty in
this derived time-lag redshift accounts for the uncertainty
in the lag and the peak flux, but not for the systematic
uncertainty resulting from the somewhat different energy
bands used to calibrate the lag-luminosity relation and for
the {\it Swift} observation.  Given the large redshift
range over which bursts are detected, the agreement between
this time-lag redshift and the spectroscopic redshift is
impressive.

The UVOT lightcurves are consistent with the existence of an early reverse shock component  
(M\'esz\'aros \& Rees 1997; Sari \& Piran 1999). The best-fit initial decay index is sensitive to whether we
include a re-brightening episode in our model, as suggested by Klotz et al. (2005). It ranges from $\alpha_1 \sim -1.5$ 
without re-brightening to $\alpha_1 \sim -2.1$ with re-brightening. This is in the range observed for
previous suggested detections of the reverse shock  in GRB 990123 (Akerlof 1999) and GRB 021211 
(Fox et al. 2003; Li et al. 2003), which are typically $\sim -1.9$. A shallower decay corresponds to a shallower electron 
index in the reverse shock region, $p=(-4 \alpha_1-1)/3 \sim 1.75$ (Zhang, Kobayashi \& M\'esz\'aros 2003). In 
contrast with Shao \& Dai (2005) who interpreted the tentative jet break as the forward shock peak, in our 
best fit, the forward shock peaks at a much earlier time, before 400~s. This corresponds to a typical 
``flattening"-type early afterglow (Zhang et al. 2003), which usually requires a magnetized central engine 
(see also Fan et al. 2002; Kumar \& Panaitescu 2003). The best fit temporal index that we derive for the forward shock 
component in the optical/UV without a re-brightening  is -0.62, flatter than the expectation 
of the simplest model (-0.9). However if we include a re-brightening episode as described above, the 
slope of the forward shock component steepens to $-1.04$, which would be more
consistent with the simple model.

\acknowledgments
The Swift programme is supported by NASA, PPARC and ASI (contract number I/R/039/04).

\end{document}